\documentclass[longauth]{aa}
\usepackage{txfonts}
\usepackage{graphicx}
\usepackage{longtable}
\usepackage{color}
\usepackage{multirow}
\usepackage{natbib,lineno}
\bibpunct{(}{)}{;}{a}{}{,}

\newcommand{\HESS}{H.E.S.S.\/ }

\begin{document}

\title{HESS~J1943+213: a candidate extreme BL Lacertae object}
\titlerunning{HESS~J1943+213: a candidate extreme BL Lacertae object}


\small{
\author{H.E.S.S. Collaboration
\and A.~Abramowski \inst{1}
\and F.~Acero \inst{2}
\and F.~Aharonian \inst{3,4,5}
\and A.G.~Akhperjanian \inst{6,5}
\and G.~Anton \inst{7}
\and A.~Balzer \inst{7}
\and A.~Barnacka \inst{8,9}
\and U.~Barres~de~Almeida \inst{10}\thanks{supported by CAPES Foundation, Ministry of Education of Brazil}
\and A.R.~Bazer-Bachi \inst{11}
% ...
}

\institute{
Universit\"at Hamburg, Institut f\"ur Experimentalphysik, Luruper Chaussee 149, D 22761 Hamburg, Germany \and
Laboratoire de Physique Th\'eorique et Astroparticules, Universit\'e Montpellier 2, CNRS/IN2P3, CC 70, Place Eug\`ene Bataillon, F-34095 Montpellier Cedex 5, France \and
Max-Planck-Institut f\"ur Kernphysik, P.O. Box 103980, D 69029 Heidelberg, Germany \and
Dublin Institute for Advanced Studies, 31 Fitzwilliam Place, Dublin 2, Ireland \and
National Academy of Sciences of the Republic of Armenia, Yerevan  \and
Yerevan Physics Institute, 2 Alikhanian Brothers St., 375036 Yerevan, Armenia \and
Universit\"at Erlangen-N\"urnberg, Physikalisches Institut, Erwin-Rommel-Str. 1, D 91058 Erlangen, Germany \and
Nicolaus Copernicus Astronomical Center, ul. Bartycka 18, 00-716 Warsaw, Poland \and
CEA Saclay, DSM/IRFU, F-91191 Gif-Sur-Yvette Cedex, France \and
University of Durham, Department of Physics, South Road, Durham DH1 3LE, U.K. \and
Centre d'Etude Spatiale des Rayonnements, CNRS/UPS, 9 av. du Colonel Roche, BP 4346, F-31029 Toulouse Cedex 4, France
% ...
}

\newpage

\authorrunning{HESS Collaboration}


\date{Released 2009 Xxxxx XX}

\abstract
{The \HESS Cherenkov telescope array has been surveying the Galactic Plane for new VHE ($>$100 GeV) gamma-ray sources.}
{We report on a newly detected point-like source, HESS~J1943+213. This source coincides with an unidentified hard X-ray source IGR~J19443+2117, which was proposed to have radio and infrared counterparts.}
{We combine new \HESS, {\em Fermi}/LAT and Nan\c{c}ay Radio Telescope observations with pre-existing non-simultaneous multi-wavelength observations of IGR~J19443+2117 and discuss the likely source associations as well as the interpretation as an active galactic nucleus, a gamma-ray binary or a pulsar wind nebula.}
{HESS~J1943+213 is detected at the significance level of $7.9 \sigma$ (post-trials) at
RA(J2000) $=19^{\rm h} 43^{\rm m} 55^{\rm s} \pm 1^{\rm s}_{\rm stat} \pm 1^{\rm s}_{\rm sys}$,
DEC(J2000) $= +21^{\circ} 18' 8'' \pm 17''_{\rm stat} \pm 20''_{\rm sys}$. 
The source has a soft spectrum with photon index
$\Gamma = 3.1 \pm 0.3_{\rm stat} \pm 0.2_{\rm sys}$
and a flux above 470 GeV of $(1.3 \pm 0.2_{\rm stat} \pm 0.3_{\rm sys}) \times 10^{-12}$ cm$^{-2}$~s$^{-1}$.
There is no {\em Fermi}/LAT counterpart down to a flux limit of $6 \times 10^{-9}$ cm$^{-2}$~s$^{-1}$ in the 0.1--100 GeV energy range (95\% confidence upper limit calculated for an assumed power-law model with a photon index $\Gamma=2.0$). The data from radio to VHE gamma-rays do not show any significant variability.}
{The lack of a massive stellar counterpart disfavors the binary hypothesis, while the soft VHE spectrum would be very unusual in case of a pulsar wind nebula. In addition, the distance estimates for Galactic counterparts places them outside of the Milky Way. All available observations favor an interpretation as an extreme, high-frequency peaked BL Lac object with a redshift $z>0.14$. This would be the first time a blazar is detected serendipitously from ground-based VHE observations, and the first VHE AGN detected in the Galactic Plane.}

 
\maketitle
%

\section{Introduction}\label{intro}
The High Energy Stereoscopic System (\HESS) is an array of four imaging atmospheric Cherenkov telescopes situated in the Khomas Highland of Namibia \citep{hess-crab}.  H.E.S.S. collaboration has been conducting a systematic scan of the Galactic Plane, increasing the survey sensitivity and expanding the catalog of very high energy  (VHE, $>$100 GeV) sources with new intrinsically faint or distant objects. In the survey, which now covers most of the Galactic Plane as seen from the Southern Hemisphere, a rich population of VHE galactic sources has been discovered \citep{aharonian2005b,hessgalaxy, chaves2009}. 

The majority of known VHE sources are extended beyond the $\sim 0.1^{\circ}$ \HESS point spread function (PSF). The few sources in the Galactic Plane that appear point-like are often associated with VHE gamma-ray emitting high-mass X-ray binaries (HMXB) which include the very well established binaries \object{PSR B1259-63} \citep{hesspsrb1259}, \object{LS 5039} \citep{hessls5039} and \object{LS I +61 303} \citep{magiclsi61303}. The point-like VHE source \object{HESS J0632+057} is now a strong candidate for a HMXB system following a recent multi-wavelength campaign \citep{discoveryhessj0632,hinton2009}.
In addition several young pulsar wind nebulae (PWN), including the \object{Crab} Nebula \citep{hess-crab}, are also unresolved by \HESS. Outside of the Galactic Plane the point-like sources are associated with active galactic nuclei (AGN). More than 30 AGN, mostly BL Lacertae (BL Lac) objects, have been detected in VHE gamma-rays up to now\footnote{Based on the online catalog for TeV Astronomy TeVCat located at http://tevcat.uchicago.edu, provided by Scott Wakely \& Deirdre Horan.}. None have been detected yet serendipitously in the Galactic Plane survey.

Here, we present the discovery of a new unresolved VHE gamma-ray source, HESS~J1943+213, located in the Galactic Plane and spatially coincident with an unidentified INTEGRAL source, IGR~J19443+2117 \citep{bird2007}. This paper is organized as follows. In Sect. \ref{s:hessanalysis}, we present the \HESS data analysis and results. In Sect. \ref{s:counterparts}, we present the sources found within the \HESS source error circle and evaluate the plausibility of each of them being a counterpart of the \HESS source. This section also includes an analysis of archival INTEGRAL and {\em Fermi} data, as well as of data taken recently by the Nan\c{c}ay Radio Telescope. Assuming that all these sources are counterparts to the H.E.S.S. source, Sect. \ref{s:discussion} discusses the possible association of HESS~J1943+213 with an HMXB, PWN and AGN. The conclusions are set out in Sect.\ref{s:conclusions}.


\section{\HESS data}
\label{s:hessanalysis}
The source was initially discovered in the Galactic Plane scan data collected between 2005 and 2008. Dedicated observations were performed between May and August 2009. These observations were taken in {\it wobble} mode, where the telescopes point in a direction typically at an offset of $0.5^{\circ}$ from the nominal target position. The offset angle of the scan-mode observations varies between $0.5^{\circ}$ and $2^{\circ}$. After applying selection cuts to the data, to reject periods affected by poor weather conditions and hardware problems, the total observation time used for analysis amounts to 38.4 hours. This is equivalent to live-time of a 24.8 hours on axis due to reduced acceptance for runs (individual continuous observation segments of $\sim28$ minutes) taken with large offsets from the nominal source position. The mean zenith angle of the observations is $49^{\circ}$. 

\begin{figure}
\begin{center}
\includegraphics[width=0.95\linewidth]{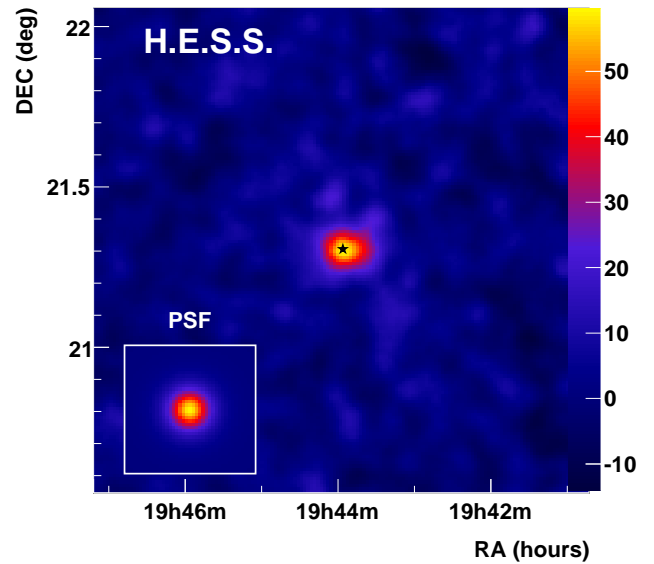}
\caption{Excess map of the field around the position of HESS~J1943+213, smoothed by the PSF of the instrument (68\% containment radius of 0.064 deg).}
\label{fig:map}
\end{center}
\end{figure}

\begin{figure}
\begin{center}
\includegraphics[width=0.95\linewidth]{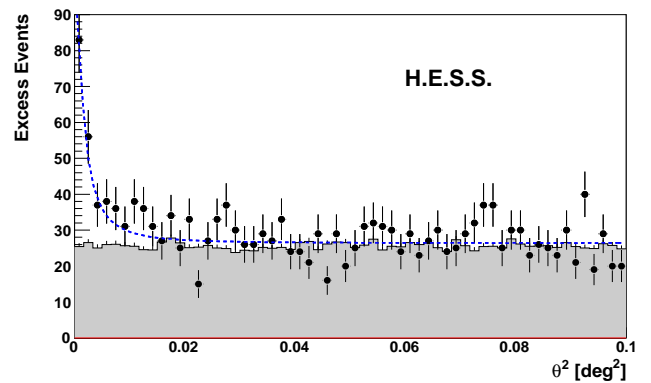}
\caption{ Distribution of squared angular distance ($\theta^2$) for gamma-like events and
normalized background events. The dashed line denotes a point source profile.}
\label{fig:theta}
\end{center}
\end{figure}

\begin{figure}
\begin{center}
\includegraphics[width=0.95\linewidth]{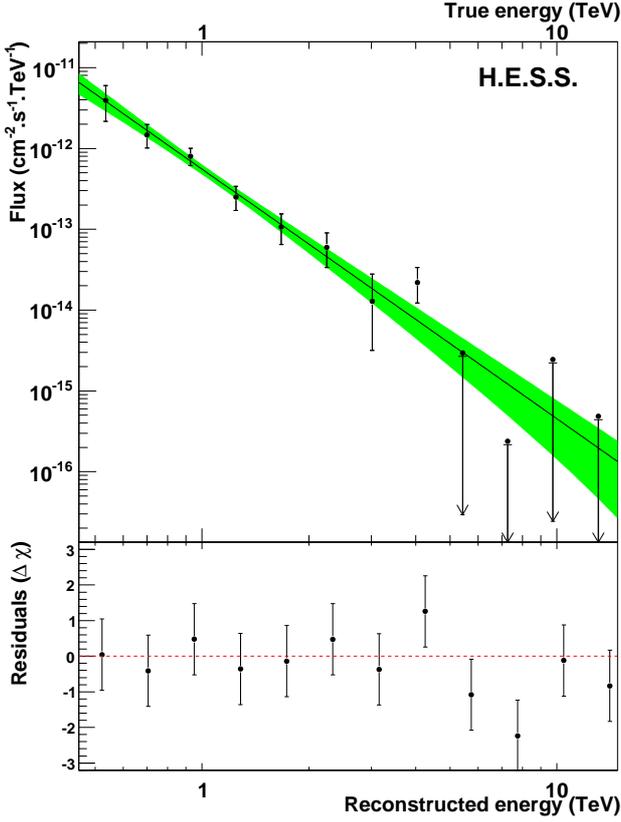}
\caption{Time-averaged VHE spectrum observed from HESS~J1943+213. The shaded area represents 
the $1\sigma$ confidence level of the fitted spectrum using a power-law hypothesis. Only the statistical 
errors are shown and the upper limits are 68\% confidence level. The lower panel shows the residual 
of the spectrum fit, which are the difference, in each reconstructed energy bin, between expected
 and observed number of excess events, normalized to uncertainty on the latter ($(N_{obs} -  N_{exp})/\Delta N_{exp}$).}
\label{fig:spectrum}
\end{center}
\end{figure}

The data were calibrated according to \cite{hess-calibration}. Energies are reconstructed taking the effective optical efficiency into account \citep{hess-crab}. Event reconstruction and separation of gamma-ray-like events from cosmic-ray background were made using the {\it Model Analysis} and standard cuts \citep{denaurois2009}, in which shower images of all triggered telescopes are compared to a pre-calculated model by means of a log-likelihood minimization.

For the spectral analysis, on-source data are taken from a circular region of radius $\theta^2=0.01$ deg$^2$ around the source position, and the background is subtracted using the event background rate estimated with theâ reflected backgroundâ model \citep{hess-crab} for which several off regions at the same distance to the center of the camera as the target position are used (excluding the region close to the source).

A total of $N_\mathrm{On}=281$ on-sources events and $N_\mathrm{Off} = 4086$ off-source events are measured. The on-off normalization factor is $\alpha = \Omega_\mathrm{On}/\Omega_\mathrm{Off} = 0.0379$, where $\Omega$ is the solid angle of the respective on- and off-source regions. The observed excess is $N_\gamma = N_\mathrm{On} -  \alpha N_\mathrm{Off} = 126$ $\gamma$-rays, corresponding to a significance of $8.9\sigma$ pre-trials standard deviations according to Equation 6 from \cite{liandma}.

In Fig. \ref{fig:map} we plot an excess map of the sky around the HESS~J1943+213 position  smoothed by the PSF of the instrument. The angular distribution of events shown in Fig. \ref{fig:theta} is consistent with that of a point source. After correcting for trials \citep{hessgalaxy} the statistical significance of the detection is $7.9\sigma$.  A fit of a point-like source convolved with the H.E.S.S. PSF of the excess yields a position at the 68\% confidence level, RA(J2000) $ =19^{\rm h} 43^{\rm m} 55^{\rm s} \pm 1^{\rm s}_{\rm stat} \pm 1^{\rm s}_{\rm sys}$, DEC(J2000) $ = +21^{\circ} 18' 8'' \pm 17''_{\rm stat} \pm 20''_{\rm sys}$.  The source is located in the Galactic plane at $l = 57.76^{\circ}$ and $b=-1.29^{\circ}$. The intrinsic source extension, calculated at the best fit position is smaller than $2.8'$ at $3\sigma$ confidence level.

Figure \ref{fig:spectrum} shows the time-averaged differential spectrum which was derived using the forward folding technique with an energy threshold of 470 GeV. Between 470 GeV and $\sim6$ TeV this spectrum is well described (Log-likelihood chance probability 45\%) by a 
power-law $dN/dE = \Phi_0(E/1{\rm TeV})^{-\Gamma}$ with photon index $\Gamma = 3.1 \pm 0.3_{\rm stat} \pm 0.2_{\rm sys}$ and normalization at 1 TeV $\Phi_0({\rm 1 TeV}) = (5.6 \pm 0.8_{\rm stat} \pm 1.1_{\rm sys}) \times 10^{-13}$ cm$^{-2}$~s$^{-1}$ TeV$^{-1}$. The 68\% confidence level upper limits for the four highest-energy bins shown in Fig. \ref{fig:spectrum} were calculated using the method of \cite{feldman1998}. The integrated flux above 470 GeV is $I(>470 {\rm GeV}) = (1.3 \pm 0.2_{\rm stat} \pm 0.3_{\rm sys}) \times 10^{-12}$ cm$^{-2}$~s$^{-1}$, 
which corresponds to $\sim 2\%$ of the Crab Nebula flux above the same energy threshold \citep{hess-crab}. 
{In Fig. \ref{fig:periodogram} we show a Lomb-Scargle periodogram for HESS~J1943+213 and in Fig. \ref{fig:lightcurve} lightcurves. There is no significant variability in the data with chance probabilities of the null hypothesis of steady emission respectively of 10.4\% and 6.2\% for the night by night and month by month lightcurve (using statistical uncertainties only).}

All the \HESS results presented in this paper have been cross-checked with an independent calibration procedure and a different background discrimination method \citep{ohm2009}. 

\begin{figure}
\begin{center}
\includegraphics[width=0.95\linewidth]{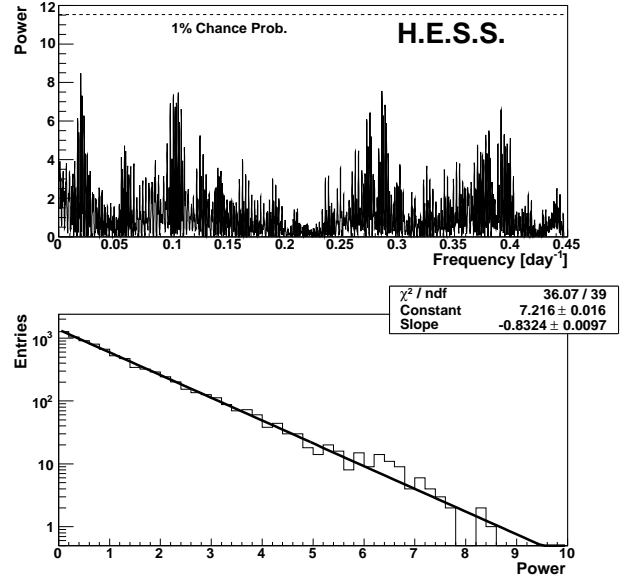}
\caption{top: Lomb-Scargle periodogram of HESS~J1943+213 based on the run-by-run lightcurve. The dashed line corresponds to the 1\% chance probability level. Bottom: distribution of power, adjusted by an exponential distribution. In the case of a pure white noise, a slope of -1 is expected.}
\label{fig:periodogram}
\end{center}
\end{figure}

\begin{figure}
\begin{center}
\includegraphics[width=0.95\linewidth]{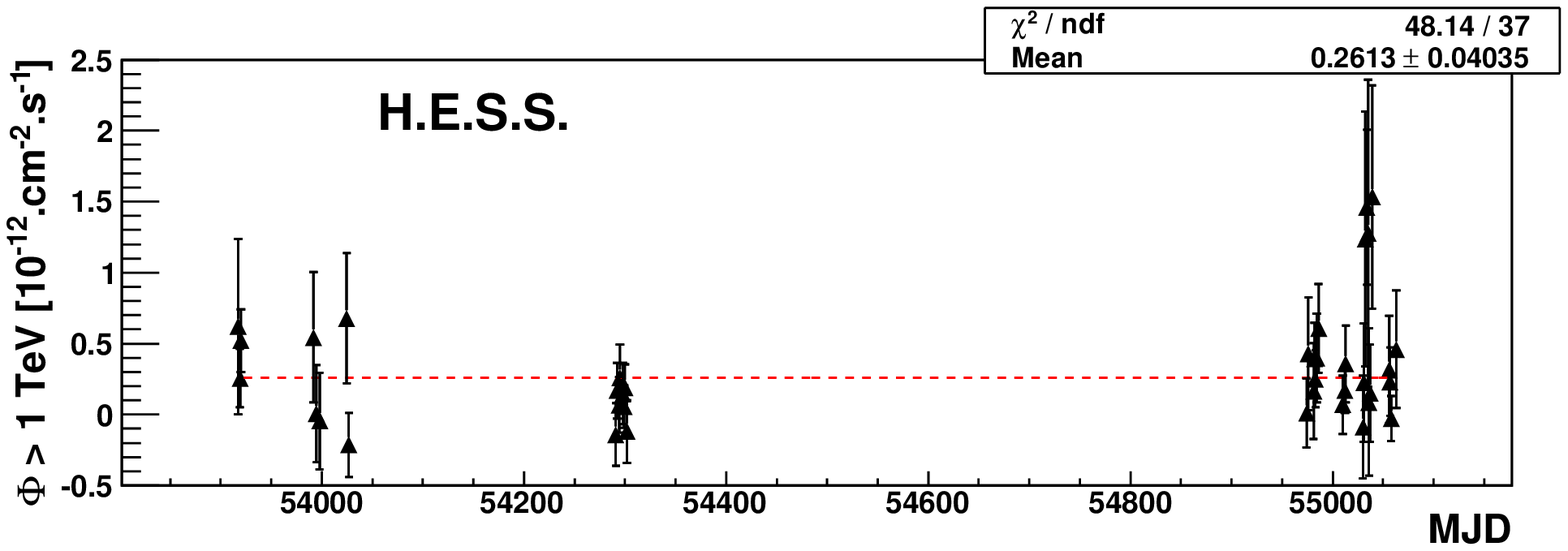}
\includegraphics[width=0.95\linewidth]{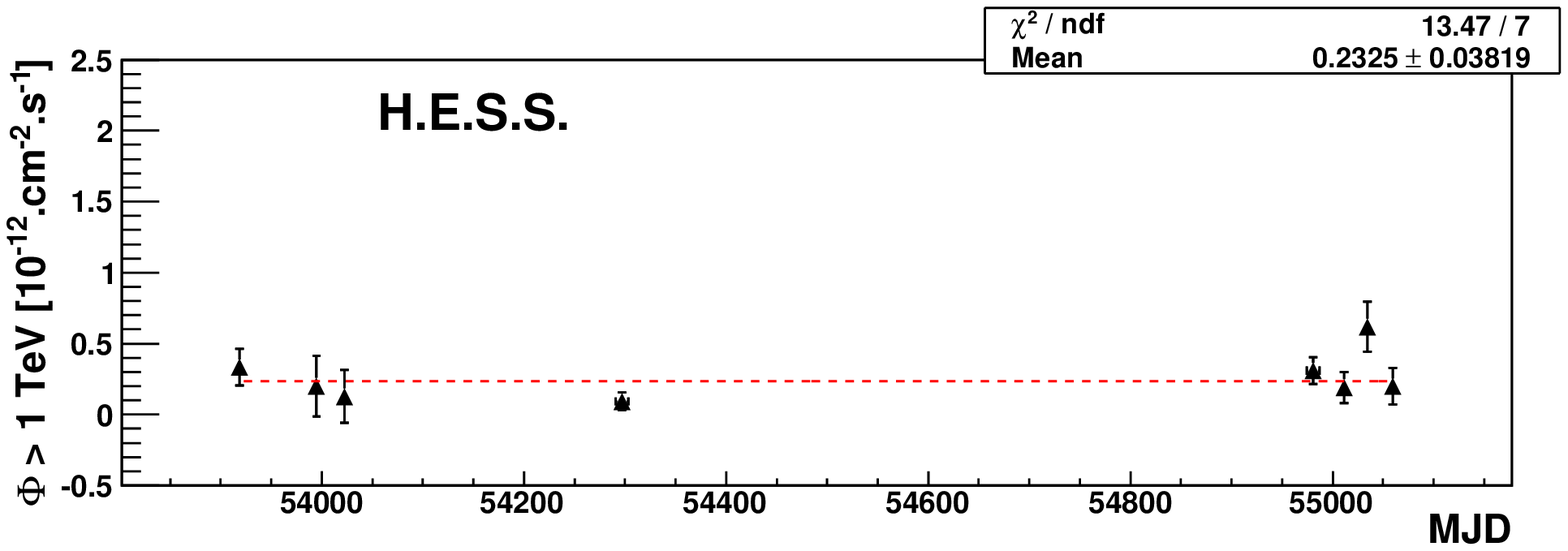}
\caption{Integral flux $I(>1 {\rm TeV})$ of HESS~J1943+217, measured by \HESS during each observing night (top) and averaged per month (bottom). 
Only the statistical errors are shown. The dashed line corresponds in each case to the best fit to a constant level.}
\label{fig:lightcurve}
\end{center}
\end{figure}

\begin{figure}
\begin{center}
\includegraphics[width=0.95\linewidth]{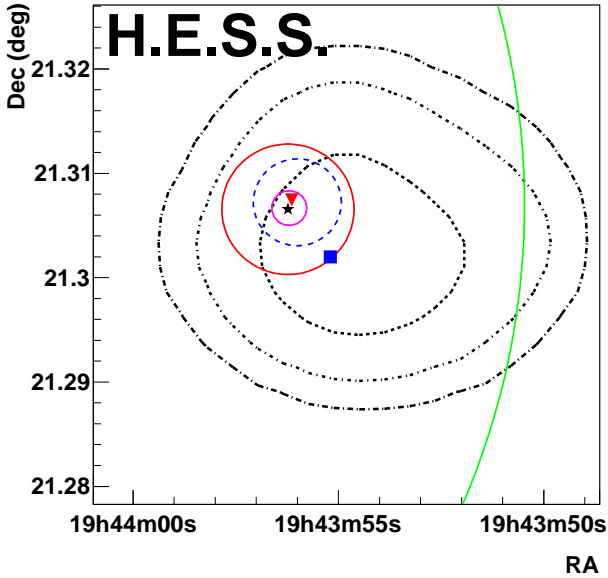}
\caption{The dot-dashed lines correspond to (from the innermost) 68\%, 95\% and 99\% best fit confidence 
level contours for the HESS~J1943+213. The positions of the possible counterparts are indicated: 
the partial circle in green/light grey is the (large) INTEGRAL error circle for \object{IGR~ 19443+2117}; 
the red/grey solid circle is radio beam size with NVSS J194356+211826 in center; 
the blue/grey dashed circle is the NRT position;
the small magenta/grey solid circle is the ROSAT 90\% position circle with 1RXH J194356.2+211824 in center;
the triangle is used for NVSS position;
the blue square is used for BAT position;
the star is used for both CXOU~J194356.2+211823 and 2MASS~J19435624+2118233.}
\label{fig:counterparts}
\end{center}
\end{figure}



\begin{table*}
\caption{The multi-wavelength fluxes and flux upper limits of potential counterparts of HESS~J1943+213.} 
\label{table1}
\begin{center}
\begin{tabular}{l l l l l l l l l }
\hline\hline

Observatory & Band & RA (J2000) & DEC (J2000) & Position & $\theta\ ^{(9)}$ & $\alpha=\Gamma-1$ & Flux & Ref. \\
 &  & [hh mm ss.s] & [$^{\circ}$ $^{\prime}$ $^{\prime\prime}$] & Error 90\% & [$''$] &   & [erg cm$^{-2}$ s$^{-1}$] &   \\

\hline

NVSS$^{(1)}$ & 1.4 GHz & 19 43 56.14 & +21 18 26.9 & $0.45''$, $0.56''$ & 24.7 & $0.32\pm0.02^{\ (10)}$ & $(1.44 \pm 0.05) \times 10^{-15\ (12)}$ & a, b \\
NRT$^{(1)}$ & 1.4 GHz & 19 43 56 & +21 18 26 & $15''\ ^{(8)}$ & 22.8 & $0.6\pm0.2$ & $(1.55 \pm 0.02) \times 10^{-15\ (12)}$ \\
2MASS$^{(2)}$ & $2.2\ \mu$m (K) & 19 43 56.244 & +21 18 23.38 & $<0.2''$ & 23.2 & -- & $2.37 \times 10^{-12\ (12, 13)}$ & c \\
Swift/UVOT & $0.54\ \mu$m (V) & 19 43 56.20 & +21 18 22.95 & -- & 22.5 & -- & $<1.6 \times 10^{-13\ (12, 13)}$ & g \\
ROSAT/HRI$^{(3)}$ & 0.1--2 keV & 19 43 56.2 & +21 18 24 & $6''$ & 23.2 & --\ $^{(14)}$ & $(3.82 \pm 0.55) \times 10^{-11\ (14)}$ & d \\
{\em Chandra}$^{(4)}$ & 0.3--10 keV & 19 43 56.23 & +21 18 23.6 & $0.64''$ & 23.2 & $0.83\pm0.11$ & $(2.9^{+2.4}_{-0.5}) \times 10^{-11}$ & e \\
Swift/XRT$^{(5)}$ & 2--10 keV & 19 43 56.20 & +21 18 22.95 & $3.53''$ & 22.5 & $1.04\pm0.12$ & $(1.83\pm0.04) \times 10^{-11}$ & f, g \\
Swift/BAT$^{(6)}$ & 15--150 keV & 19 43 54.35 & +21 18 08.2 & $2.88'$ & 9.1 & --\ $^{(11)}$ & $(1.8 \pm 0.2) \times 10^{-11}$ & h \\
Swift/BAT$^{(5)}$ & 14--195 keV & 19 43 55.2 & +21 18 07.2 & -- & 2.9 & $1.1^{+0.3}_{-0.2}$ & $(2.9^{+0.5}_{-0.5}) \times 10^{-11}$ & k \\
INTEGRAL/IBIS$^{(7)}$ & 20--100 keV & 19 44 09.36 & +21 18 25.2 & $4.4'$ & 201 & $1.12\pm0.22$ & $(1.12\pm0.22) \times 10^{-11}$ & g, i \\
INTEGRAL/SPI$^{(7)}$ & 100--200 keV & 19 44 23.13 & +21 16 36.4 & -- & 404 & -- & $<2.6 \times 10^{-11}$ & j \\
{\em Fermi}/LAT & 0.1--100 GeV & 19 43 56.23 & +21 18 23.6 & -- & 23.2 & $1.0^{\ (15)}$ & $<7 \times 10^{-12}$ \\
{\em Fermi}/LAT &  &  &  &  &  & $0.5^{\ (15)}$ & $<1 \times 10^{-12}$ \\
\HESS & $>470$ GeV & 19 43 55 & +21 18 8 & $34''_{\rm stat}$, $21''_{\rm sys}$ & 0 & $2.1 \pm 0.3_{\rm stat}\pm0.2_{\rm sys}$ & $(1.8^{+0.6_{\rm stat}+0.5_{\rm sys}}_{-0.4_{\rm stat}-0.4_{\rm sys}}) \times 10^{-12}$ \\

\hline
\end{tabular}
\end{center}
{\bf Notes.}
(1) NVSS J194356+211826;
(2) 2MASS~J19435624+2118233;
(3) 1RXH J194356.2+211824;
(4) CXOU~J194356.2+211823;
(5) SWIFT~J1943.5+2120;
(6) PBC J1943.9+2118;
(7) IGR~J19443+2117;
(8) Error for right ascension only;
(9) Angular distance from the centroid of \HESS source position;
(10) Spectral index given for all combined radio measurements between 327 and 4850 MHz, listed in SPECFIND V2.0 \citep{vollmer2009};
(11) Hardness ratio Rate(30--150 keV)/Rate(14--30 keV) = $1.0 \pm 0.4$;
(12) $EF(E)$;
(13) The flux is not corrected for extinction;
(14) Unabsorbed flux obtained from ROSAT/HRI count-rate using a Mission Count Rate Simulator {\tt PIMMS}, assuming $N_{\rm H}$ and $\alpha_{\rm X}$ identical with {\em Chandra}. The flux error includes uncertainties of  $N_{\rm H}$ and $\alpha_{\rm X}$;
(15) The spectral index was assumed for upper limit calculation.

{\bf References.}
(a) \cite{condon1998};
(b) \cite{vollmer2009};
(c) \cite{skrutskie2006};
(d) \cite{Rosat};
(e) \cite{tomsick2009};
(f) \cite{malizia2007};
(g) \cite{landi2009};
(h) \cite{cusumano2010};
(i) \cite{bird2010};
(j) \cite{bouchet2008};
(k)  \cite{baumgartner2011}.

\end{table*}

\section{Counterparts}
\label{s:counterparts}

We searched for possible H.E.S.S. source counterparts in a number of radio, IR, optical and X-ray catalogs. Table \ref{table1} contains the position and spectral information with uncertainties for each source discussed in this section by reason of finding them plausible counterpart of HESS~J1943+213. Fig. \ref{fig:counterparts} shows the \HESS source position confidence contours along with the candidate counterparts positions.

\subsection{X-ray counterpart}
The \HESS source is located within the error circle ($4.4'$) of an unidentified hard X-ray INTEGRAL/IBIS source IGR~J19443+2117, the centroid position of which is about 3.3 arcmin away from the \HESS position. A counterpart to IGR~J19443+2117 was detected in the soft X-ray domain by {\em Chandra} (CXOU J194356.2+211823; \citealt{tomsick2009}) and Swift (SWIFT J1943.5+2120; \citealt{landi2009}) with exposure times of 4.8 ks and 11 ks, respectively. 
The same X-ray source was detected in the past by ROSAT/HRI (1RXH J194356.2+211824, exposure 1.3 ks, \citealt{Rosat}) 

The combined power-law fit to Swift/XRT and INTEGRAL/IBIS data gives a spectral index $\alpha_{\rm X} = 1.04^($\footnote{$\alpha=\Gamma-1$; in the following, the subscript of spectral index $\alpha$ denotes a single waveband with R for radio and X for X-ray; or a broad waveband with RX for a band extending from radio to X-ray, RO from radio to optical and OX from optical to X-ray.}$^)$
 and fluxes $F_{\rm 2-10 keV}=1.83~\times~10^{-11}$ erg~cm$^{-2}$~s$^{-1}$, $F_{\rm 20-100 keV} = 1.12 \times 10^{-11}$ erg~cm$^{-2}$~s$^{-1}$ \citep{landi2009}. Within the error of the cross-calibration constant between XRT and IBIS equal $0.6$, the IBIS data is in agreement with that of XRT.

The {\em Chandra} observation (performed two years later in 2008), is consistent with the Swift results and gives $F_{\rm 0.3-10 keV} = 2.9 \times 10^{-11}$ erg~cm$^{-2}$~s$^{-1}$ with a spectral index $\alpha_{\rm X}=0.83$ \citep{tomsick2009}. The oldest soft X-ray observation of the source comes from ROSAT/HRI. The Mission Count Rate Simulator {\tt PIMMS} was used in order to obtain the flux from the HRI count-rate. Assuming a hydrogen column density (see Sect. \ref{IRcounterpart}) and spectral index identical to the {\em Chandra} source, the flux equals $F_{0.1-2 {\rm keV}} = 3.82 \times 10^{-11}$ erg cm$^{-2}$ s$^{-1}$, this is in agreement with both {\em Chandra} and Swift fluxes within errors.

The INTEGRAL/SPI observations provide an upper limit on the source flux above 100 keV of $F_{\rm >100 keV} < 2.6 \times 10^{-11}$ erg~cm$^{-2}$~s$^{-1}$ \citep{bouchet2008}. An X-ray source consistent with all above X-ray observations is also present 
in the Swift/BAT hard X-ray Catalog (PBC J1943.9+2118, \citealt{cusumano2010}) with $F_{\rm 14 -150 keV} = 1.8 \times 10^{-11}$ erg~cm$^{-2}$~s$^{-1}$, and 
in the 58 months Swift/BAT catalog \citep{baumgartner2011} with $F_{\rm 14 -195 keV} = 2.86 \times 10^{-11}$ erg~cm$^{-2}$~s$^{-1}$ and $\alpha_{\rm X} = 1.08$. A power-law model fit to the spectrum provided by \cite{baumgartner2011}\footnote{http://heasarc.gsfc.nasa.gov/docs/swift/results/bs58mon/} yields $\chi^2/$dof = 2.4/6, whereas a cut-off power-law model fit yields $\chi^2/$dof = 3/5. There is no evidence for a cut-off in the BAT spectrum.

The public archival INTEGRAL data were analyzed to obtain a lightcurve. The IBIS lightcurve was constructed in the 18--60 keV energy band, and with individual points corresponding to about $\sim2000$ s (INTEGRAL science window). We used the same methods as those used in the construction of the 4$^{th}$ IBIS Catalog \citep{bird2010}. Individual science windows that had exposures of less than 500 s or where the source was at a large off-axis angle ($>12^{\circ}$ where the instrument flux calibration is not optimized) were excluded from the lightcurve. This produced a lightcurve of 1295 observations, spanning MJD 52704.16--54604.20, with a cumulative exposure of $\sim$2.6 Ms. The source is detected at 5.3$\sigma$ in the 18--60 keV band with a weighted mean flux of $0.6~\pm$~0.1 mCrab; consistent with the reported catalog flux of \cite{bird2010}.  The source is weak and its lightcurve is compatible with white noise. Fitting the constant to the lightcurve yields $\chi^2/$dof= 1.2 for 1293 dof.

Based on the facts that the ROSAT/HRI, {\em Chandra}, Swift/XRT and INTEGRAL/IBIS fluxes (within the error range of the cross-calibration constant) are constant and consistent with each other, that the ROSAT, {\em Chandra} and Swift sources are only $23''$ away from the centroid of the \HESS source position, and that there are no other identified X-ray sources in the \HESS error circle, we conclude that these X-ray observatories detected the same object and that this object is a very likely counterpart to the \HESS source. Assuming this association is correct, in the following, we adopt the {\em Chandra} source location with its uncertainty for evaluation of potential counterparts in IR and radio bands.

\subsection{IR, optical and UV counterpart}
\label{IRcounterpart}
The 2MASS Catalog lists 19 infrared sources within the 90\% \HESS error circle, however only one faint unidentified source 2MASS J19435624+2118233 is situated within the small {\em Chandra} error circle \citep{skrutskie2006,landi2009,tomsick2009}. The angular separation between the {\em Chandra} position and the 2MASS object is $0.3''$. The source is viewed across the Galactic Plane, therefore, its IR and optical fluxes are strongly affected by extinction. The magnitude lower limits in the J and H bands are 14.53 and 13.36, respectively, and the source magnitude in the K band is 13.98. The lower limit on optical and UV magnitudes of the source from Swift/UVOT observations are $V > 20.3$ and $UVM2 > 21.5$ \citep{landi2009}. Based on maps of Galactic dust infrared emission of \cite{schlegel1998}, the extinction coefficients in the direction of the source are $A_{\rm V} = 8.96$ mag, $A_{\rm J} = 2.49$ mag, $A_{\rm H} = 1.59$ mag, $A_{\rm K} = 1.02$ mag for a color excess $E(B-V) = 2.81 \pm 0.25$. The extinction corrected magnitudes are $V_0 > 11.34$, $J_0 > 12.04$, $H_0 > 11.77$, $K_0 = 12.96$. The extinction correction of the UV magnitude, based on extinction curves of \cite{cardelli1989}, gives $UVM2_0 > 26.04$ mag.

The measured color excess corresponds to a hydrogen column density $N_{\rm H}=(1.53 \pm 0.02) \times 10^{22}$ cm$^{-2}$ \citep{predehl1995}, which is higher than the $N_{\rm H}$ measured by the survey of Galactic neutral hydrogen, $(8.4 \pm 1.3) \times 10^{21}$ cm$^{-2}$ \citep{kalberla2005}. The $N_{\rm H}$ obtained from X-ray {\em Chandra} and Swift spectra are $(1.89^{+0.25}_{-0.22}) \times 10^{22}$ cm$^{-2}$ and $(1.37^{+0.12}_{-0.13}) \times 10^{22}$ cm$^{-2}$ respectively \citep{tomsick2009, landi2009}, thus consistent with the column density obtained from dust maps. Therefore, there is no conclusive evidence that the flux of the X-ray source is intrinsically absorbed (in spite of observed $N_{\rm H}$ being in excess of Galactic neutral hydrogen measurements). However, dust maps used to calculate extinction may not be precise at Galactic latitudes lower than $5^{\circ}$ \citep{schlegel1998}.

The potential IR counterpart to IGR~J19443+2117 was observed at ESO by Masetti et al. (private communication, publication in preparation), during a campaign aimed at identifying INTEGRAL sources. A {\em preliminary} IR spectrum of 2MASS J19435624+2118233 obtained with SofI spectro-imager at NTT \citep{moorwood1998} is heavily reddened and does not show any readily apparent emission lines.

\subsection{Radio counterpart}
There is only one radio source in the NVSS Catalog within the \HESS error circle (NVSS J194356+211826; \citealt{condon1998}). It is located $24.7''$ away from the best fit \HESS source position and $3.5''$ from the {\em Chandra} source position. Although the {\em Chandra} source is outside of the NVSS error circle ($\sim 0.5''$), it is still well within the radio PSF of $45''$. The radio source has been detected by various survey programs between 327 and 4850 MHz at positions which differ from NVSS by $1.7''-40''$ \citep{taylor1996, gregory1996, reich1984, furst1990, becker1991, gregory1991, griffith1991}. The spectral index calculated for nine combined measurements is $\alpha_{\rm R}=0.32$ \citep{vollmer2009} and the source flux density measured by NVSS at 1400 MHz is $102.6$ mJy \citep{condon1998}. There is no evidence for flux variation over the 12-year time span of the collected data.

Between March and May 2010 the radio source was monitored by the Nan\c{c}ay Radio Telescope (NRT), a meridian transit telescope with a main spherical mirror of 300 m $\times$ 35 m \citep{theureau2007}. Thirteen observations were performed at two frequencies 1419 and 2392 MHz. The data processing consisted of selecting the best drift scans, background subtraction, straightening of the baseline by polynomial fit and application of a median smoothing. The calibration was corrected by the observations of the radio sources 3C123, 3C196, 3C317 and PKS 0237-23 with the same setup as the main target.

The average statistical error for all observations is $\sigma_{\rm stat} = 4$ mJy. The systematic errors $\sigma_{\rm sys} = 17$ mJy at low frequency and $10$ mJy at high frequency, take into account the uncertainty in the background subtraction and baseline contamination by secondary beams. The former mostly affects the low frequency observations due to larger beam size (more background sources are found within the beam together with the source of interest). The 3-months lightcurve is constant within the range of systematic errors. The constant fit to the data yields $\chi^2/$dof$ < 1$ at both frequencies. The average fluxes at low and high frequency are $111\pm21$ mJy and $86\pm14$ mJy, respectively. The average spectral index is $\alpha_{\rm R} = 0.59 \pm 0.16$. The low frequency flux is in agreement with NVSS but the spectral index is somewhat higher than the archival observations.

\begin{figure*}
\begin{center}
\includegraphics[width=0.95\linewidth]{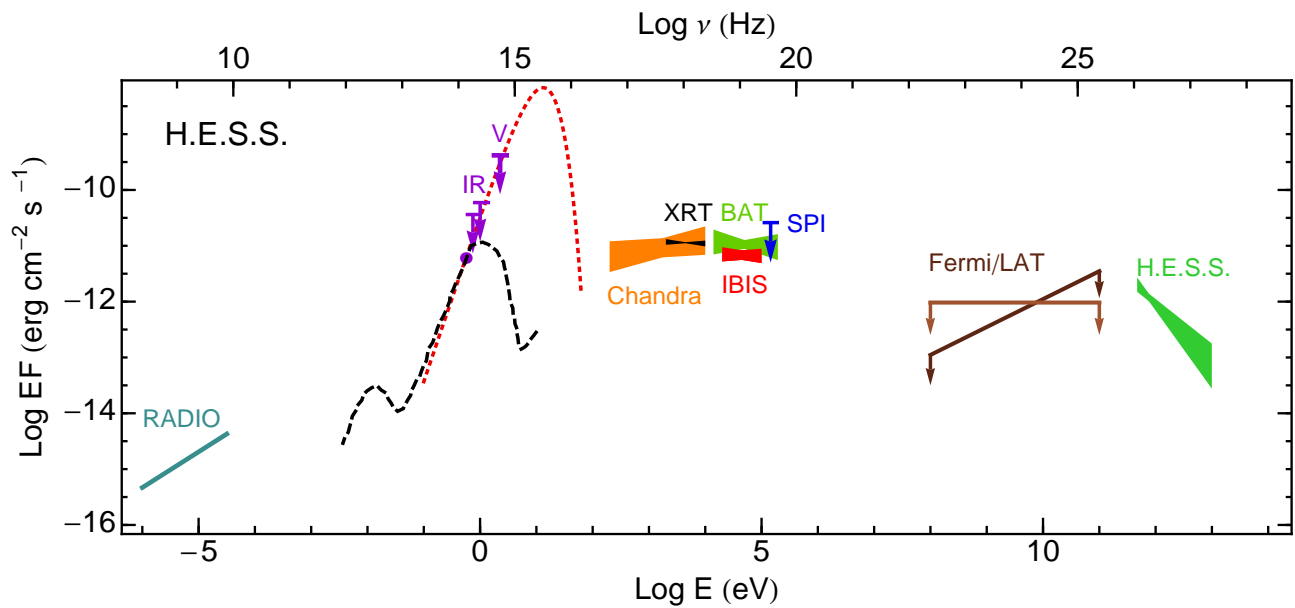}
\caption{Spectral energy distribution of the sources discussed in Sect.\ref{s:counterparts}. The details of the data and sources used in this plot are given in Table \ref{table1} and in Sect. \ref{IRcounterpart}. The dashed black line is the spectral template of elliptical galaxies from \cite{silva1998} at a distance of 570 Mpc. The dotted (red) line is the blackbody spectrum of a massive star with parameters identical with the companion star in the gamma-ray binary LS 5039 at a distance of 25 kpc. The data in IR (2MASS) and visual (Swift/UVOT) bands were corrected for extinction. For {\em Fermi}/LAT we plot both upper limits for $\Gamma=1.5$ (dark brown) and $\Gamma=2.0$ (light brown).}
\label{fig:sed}
\end{center}
\end{figure*}

\subsection{Fermi/LAT data}
\label{s:fermi}
The primary instrument onboard {\em Fermi}, the Large Area Telescope (LAT) is
an electron-positron pair production telescope, featuring solid state
silicon trackers and caesium iodide calorimeters, sensitive to high-energy (HE) gamma-ray photons from $\sim 20$ MeV to $> 300$ GeV \citep{atwood2009}. The data were nominally taken in survey mode; the observatory is rocked north and south on alternate orbits so that every part of the sky is observed for $\sim 30$ minutes every 3
hours.

The LAT analysis dataset spanned MJD 54682.7 - 55319.4, corresponding to a period of $\sim 21$ months. The data were reduced and analyzed using the {\em Fermi} Science Tools {\tt v9r15} package. The standard onboard filtering, event reconstruction, and
classification were applied to the data \citep{atwood2009}, and for this analysis the high-quality ("diffuse") photon event class is used. Throughout the analysis, the "Pass 6 v3 Diffuse" ({\tt P6\_V3\_DIFFUSE}) instrument response functions are applied.

Events between 0.1--100 GeV from a 10 degree region of interest centred
on the {\em Chandra} X-ray counterpart to HESS~J1943+213 were extracted and a
maximum likelihood analysis was performed on the data using the {\tt gtlike}
tool. A model was constructed containing all the point sources
within $15^{\circ}$ that are listed in the {\em Fermi}/LAT 1FGL Catalog \citep{fermicatalog}; the
point sources were modelled by a single power-law function and their
parameters were frozen to those reported in the 1FGL Catalog unless they
were within $10^{\circ}$ of HESS~J1943+213. Also incorporated were the currently recommended models for the Galactic diffuse emission
({\tt gll\_iem\_v02.fit}) and isotropic backgrounds; the Galactic diffuse
emission is fit with a power-law scaling. As HESS~J1943+213 does not
appear in the 1FGL Catalog an additional source was inserted into the
model at the {\em Chandra} source location assuming a power-law spectrum. The
likelihood analysis reveals that no significant source is detected at
the {\em Chandra} source location i.e. at the best estimated position of the potential HESS~J1943+213 counterpart. The 95\% 0.1--100 GeV flux upper limits calculated using a Bayesian approach \citep{helene1991} are $< 1 \times 10^{-9}$ cm$^{-2}$~s$^{-1}$ for an assumed spectral index $\Gamma=1.5$ and $< 6 \times 10^{-9}$ cm$^{-2}$~s$^{-1}$  for an assumed spectral index $\Gamma=2.0$. It should be noted that this is a highly complex, crowded region in which there is a large contribution from the Galactic diffuse emission.

Both upper limits are shown in Fig. \ref{fig:sed}, where the limit for $\Gamma=1.5$ is plotted in dark brown, and $\Gamma=2.0$ in light brown. The comparison of the {\em Fermi}/LAT upper limit with the flux measured by H.E.S.S., implies a break in the HE/VHE gamma-ray spectrum located between
100~GeV and 500 GeV (or even higher if the spectrum is attenuated by the extragalactic background light), and a hard HE spectrum with $\Gamma < 2.0$.

\section{Discussion}
\label{s:discussion}
The positional coincidence between the new VHE gamma-ray source HESS~J1943+213 and a radio, IR and X-ray source together with the lack of any other obvious candidate counterpart strongly suggests an association. Figure \ref{fig:sed} shows the spectral energy distribution (SED) of HESS~J1943+213 together with the radio, 2MASS, {\em Chandra}, Swift, INTEGRAL, and {\it Fermi} data. The data are not simultaneous but no significant variability was found within any energy range, therefore, we proceed with a contemporaneous discussion of the entire dataset from radio to VHE gamma-rays.

Based on the X-ray, IR and radio properties of IGR~J19443+2117, \cite{landi2009} and \cite{tomsick2009} argued briefly in favour of its AGN nature. The detection in VHE gamma-rays helps place tighter constraints on the source nature. Because of its point-like appearance in VHE gamma-rays, the source is most likely an AGN, a gamma-ray binary or a PWN (see Sect. 1). All options are discussed below in light of the multiwavelength properties presented in Sect. \ref{s:counterparts}.

\subsection{Gamma-ray binary}
The point-like appearance of HESS~J1943+213 in VHE gamma-rays and its location in the Galactic Plane may suggest a gamma-ray binary. With a handful of known systems, knowledge of these sources as a class is still limited. Known gamma-ray binaries are all radio sources with spectral indices at GHz frequencies $\alpha_{\rm R} \sim 0-0.5$ \citep{johnston1999,Ribo1999,massi2009}. Their X-ray spectra are hard ($\alpha_{\rm X} \sim 0.4-1.0$) with no cutoffs up to MeV energies \citep[e.g.,][]{takahashi2009,smith2009,uchiyama2009}. The softest VHE spectra belong to LS~I+61~303 ($\Gamma=2.7 \pm 0.5_{\rm stat} \pm 0.2_{\rm sys}$, \citealt{lsi61303}) or PSR B1259-63 ($\Gamma=2.8\pm 0.2_{\rm stat} \pm 0.2_{\rm sys}$, \citealt{psrb1259b}). In LS~5039 the photon index changes with the orbital phase between $1.85 \pm 0.25$ and $3.09\pm 0.47$ \citep{ls5039}. The binaries LS 5039, LS I+61 303 and PSR B1259-63 have been detected by {\em Fermi}/LAT and the emission of the former two shows a cutoff at a few GeV, suggesting at least two components are required to explain their HE and VHE gamma-ray emission \citep{fermilsi61303, fermils5039,fermipsrb}. No HE counterpart has been found for HESS~J0632+057.

The radio, X-ray and VHE spectral properties of HESS~J1943+213 and its possible counterpart are compatible with what is known to date about gamma-ray binaries. The detection of orbital modulation would firmly associate HESS~J1943+213 with a binary. However the lack of modulation in any of the spectral bands does not disqualify the HMXB scenario, since for example no modulation has been detected in HESS~J0632+057. 

All known gamma-ray binaries have a bright, massive star companion (O or Be spectral type) dominating the SED in the optical/IR band. Identifying the infrared source with a massive star would give credence to the identification as a binary. Furthermore, comparing the fluxes to those of the massive star in the gamma-ray binary LS 5039 (Fig. \ref{fig:spectrum}), the IR data are consistent with a massive O star only if its distance is $> 25$ kpc, i.e. if it is located beyond the edge of our Galaxy (radius $\sim 15$~kpc) which is unlikely for a massive star. Similarly, comparing to the Be counterpart of PSR B1259-63 (the least luminous and the least hot among companions of known gamma-ray binaries) the lower limit on the distance is $>22$ kpc. A large distance would also imply a 1--10 keV X-ray luminosity of about $10^{36}$ erg~s$^{-1}$ whereas known gamma-ray binary luminosities are about $10^{33}-10^{34}$ erg~s$^{-1}$. 

Such a high X-ray luminosity can be explained if the source is accreting but, if the source is powered by a pulsar as in PSR B1259-63 and based on the observed correlations in pulsar wind nebulae \citep{mattana2009}, this would require the spindown power of the pulsar to be higher than that of the Crab: $\dot E \ga 10^{38}$ erg~s$^{-1}$. Finally, the possibility that the massive star is strongly obscured by circumstellar material, as found in several high-mass X-ray binaries detected by INTEGRAL, appears unlikely as the X-ray derived $N_{\rm H}$ column density (see Sect. \ref{IRcounterpart}) is smaller than what is typically found in these systems \cite[e.g.][]{rahoui2008}.

Despite the general similarity of the radio to X-ray SED between HESS~J1943+213 and known binaries, the lack of a plausible massive stellar counterpart in the optical/IR is the main argument which disfavors the binary hypothesis. The secondary argument against the binary hypothesis is the lack of orbital modulation in any of the bands.

\subsection{Pulsar wind nebula}
\label{PWN}
When the wind ejected from a rotation-powered pulsar is confined by the pressure of the surrounding medium (supernova remnant or compressed interstellar gas), a PWN is created. PWNe constitute an abundant class of VHE Galactic sources detected by Cherenkov telescopes. Among them a number are unresolved (e.g., \citealt{aharonian2005,djannati2008,acciari2010}), including one of the most powerful, the Crab Nebula \citep{hess-crab}.

In general, the SED of PWNe consists of a broad synchrotron component extending up to high X-ray energies ($\sim$ MeV) and an inverse Compton component at HE and VHE energies \citep{gaensler2006}. If HESS~J1943+213 is a PWN, its gamma-ray (1--30 TeV) to X-ray (2--10 keV) flux ratio of $0.04$ implies that the nebula is very young $\sim 10^3$~yr, and that it contains a pulsar with a spindown power $\dot E \sim 10^{38}$ erg~s$^{-1}$ \citep{mattana2009}. The high $\dot E$ is also supported by the X-ray photon index ($\Gamma_{\rm X} \simeq 2$) \citep{gotthelf2003}. 
 
The estimated PWN age and $\dot E$ is similar to the Crab Nebula. Assuming that the luminosity of HESS~J1943+213 is comparable to the Crab Nebula, we are able to estimate an approximate upper limit on the source distance. The low VHE flux of HESS~J1943+213 being only 1.5\% of the Crab Nebula (at 2 kpc; \citealt{trimble1973}) flux, implies a distance of $\lesssim 16$ kpc. A comparison with the remaining unresolved PWNe, which are also young but less luminous (10--15\% of the Crab Nebula luminosity), would place HESS~J1943+213 at a moderate distance of about 6 kpc.

The young age of HESS~J1943+213 in the PWN scenario is consistent with it being unresolved for \HESS. However, all the unresolved VHE PWNe are extended in X-rays \citep{weisskopf2000,ng2008,matheson2010,termim2010}, whereas HESS~J1943+213 is point-like as observed by {\em Chandra} \citep{tomsick2009}. Two unidentified \HESS sources, most likely associated with bright X-ray pulsars \citep{hessgalaxy} are embedded within very faint diffuse X-ray PWNe, these are HESS~J1837-069 (AX J1838.0-0655, \citealt{gotthelf2008}) and HESS~J1616-508 (PSR J1617-5055, \citealt{kargaltsev2009}). A 4.8 ks {\em Chandra} observation would miss such emission which could explain the point-like appearance of HESS~J1943+213 in X-rays. However, both of the unidentified \HESS sources are extended in VHE gamma-rays and both have gamma-ray to X-ray flux ratios greater than unity, much higher than in case of HESS~J1943+213. 

The hard HE spectrum implied by the upper limits in Fig. \ref{fig:sed} (see Sect. \ref{s:fermi}), would be in agreement with the HE spectrum of the Crab Nebula, where an IC component has $\Gamma=1.64$ above 1 GeV \citep{fermicrab}. Two other PWN detected by {\em Fermi}/LAT (Vela and HESS J1640-465) have soft spectra but these nebulae are older (more evolved) than the Crab, and are both extended in VHE gamma-rays. The VHE spectrum of HESS~J1943+213 is significantly softer than all known VHE PWN. The softest PWN have photon indices of $2.7\pm0.3$ and there are only a handful of them \citep{kargaltsev2010}. This fact together with the lack of obvious extended X-ray emission weaken the PWN hypothesis.

\subsection{Active galactic nucleus}
 
Blazars are AGN where a relativistic jet pointing close to the line of sight produces Doppler-boosted emission \citep{blandford1978,urry1995}. The spectral energy distribution $\nu F_{\nu}$ of blazars is characterized by two non-thermal components: one peaking in the UV to X-ray range, commonly interpreted as synchrotron emission from highly relativistic electrons accelerated in the jet, and a second peaking in the gamma-ray regime being due to inverse Compton (IC) scattering of low-energy photons. Low-energy photons are the synchrotron radiation emitted either by the same population of electrons (synchrotron self-Compton, SSC), or ambient thermal photon (external Compton). The Doppler-boosted emission from the jet dominates the SED. When blazars are arranged according to their luminosity, the peak energy of both components is found to be anti-correlated with luminosity. High-frequency peak BL Lac (HBL) objects occupy the lowest end of the blazar luminosity sequence with the two spectral components peaking in X-rays (usually in the 0.1--1 keV range) and at about $100$ GeV \citep{fossati1998}. With the IC component peaking at HE, HBL objects are the best candidates for VHE sources, and indeed the class constitutes the majority of the current population of VHE AGN.

BL Lac objects are characterized in the optical/IR band by their elliptical host galaxy and have featureless optical spectra. Therefore, while the available preliminary NTT spectrum does not give a definite answer, the apparent lack of emission lines supports a classification of HESS~J1943+213 as a BL Lac object.

If the 2MASS IR counterpart of HESS~J1943+213 is an elliptical host galaxy, its flux can be  compared with a template spectrum of the elliptical galaxies, \cite[e.g.][]{silva1998}. The measured flux in $K$ band and the flux upper limits in the $H$, $J$ and $V$ bands, corrected for extinction, are consistent with an elliptical galaxy located at a distance $> 570$ Mpc ($z \gtrsim 0.14$). This is a lower limit on the distance since non-thermal emission from the jet in a BL Lac object contributes a large fraction of the optical/IR flux. At such distances, absorption of VHE gamma rays on the extragalactic background light is expected to significantly soften the VHE spectrum. The observed \HESS spectrum is indeed quite soft (photon index of 3.2). The distance also sets the peak X-ray luminosity to be $> 10^{44}$ erg s$^{-1}$. This is relatively high for a blazar but still consistent with spectra of BL Lac objects \citep{donato2001}.

Broad band spectral indices are derived from the observations for an assumed host galaxy spectrum and compared to the typical values found in blazars. The indices are defined as $\alpha_{\rm RO} = -\log(F_{\rm O}/F_{\rm R})/5.08$, $\alpha_{\rm OX} = -\log (F_{\rm X}/F_{\rm O})/2.6$, $\alpha_{\rm RX} = -\log (F_{\rm X}/F_{\rm R})/7.68$ where $F_{\rm R}$, $F_{\rm O}$ and $F_{\rm X}$ are the differential fluxes (per unit frequency) at 5 GHz, 5000 \AA, and 1 keV respectively \citep{tananbaum1979}. For HESS~J1943+213 the indices are $\alpha_{\rm RO}=0.36$, $\alpha_{\rm OX}=0.96$ and $\alpha_{\rm RX}= 0.56$. When compared to the values of $\alpha_{\rm RO}$ and $\alpha_{\rm OX}$ measured for a large group of blazars (e.g.  Fig. 12 in \citealt{plotkin2010}), HESS~J1943+213 would be classified as a radio loud, X-ray strong HBL object. The radio loud classification of HESS~J1943+213 is also supported by the value of $\log R_{\rm X} \equiv \log(\nu L_{\rm 6 cm}/L_{\rm 2-10 keV}) = -3.7$, where the radio loud sources are those with $\log R_{\rm X}>-4.5$, as defined by \cite{terashima2003}. The calculated value of $\log R_{\rm X}$ is also similar to the average value of $-3.10$ calculated for a large sample of HBL objects, and is significantly different than the average $-1.27$ for low-frequency peaked BL Lac objects and $-0.95$ for flat spectrum radio quasars \citep{terashima2003}.

\cite{costamante2002} selected candidate VHE gamma-ray emitting BL Lac objects based on their fluxes in radio ($F_{\rm R}$) and X-ray ($F_{\rm X}$) bands measured at 5 GHz and 1 keV, respectively. BL Lac objects are expected to be detected at VHE energies if $\log \nu_{\rm R}F_{\rm R} \gtrsim -14.8$ erg cm$^{-2}$ s$^{-1}$ and $\log \nu_{\rm X}F_{\rm X} \gtrsim -11.4$ erg cm$^{-2}$ s$^{-1}$. This target selection criterion has proved quite successful \citep[e.g.][]{aharonian2008a,hess2010}. For HESS~J1943+213 the fluxes are $\log \nu_{\rm R}F_{\rm R} = -14.5$ and $\log \nu_{\rm X}F_{\rm X} = -11.1$, which places the source well within the group of VHE emitting BL Lac objects. 

The X-ray spectrum of HESS~J1943+213 is hard with no apparent cut-off up to 195 keV. 
In this respect HESS~J1943+213 resembles extreme BL Lac objects, in which the synchrotron peak energy exceeds 10~keV \citep{costamante2001}. The extreme status of HESS~J1943+213 is also supported by $\log (F_{\rm X}/F_{\rm R}) = -4.3$, which is higher than $-4.5$ as is the case for extreme HBL objects \citep{rector2003}. 

The extreme BL Lac objects are the least luminous blazars in the blazar sequence, and yet these are sources in which the dissipation of power carried by the jet is most efficient and particle acceleration mechanisms are close to their limits \citep{ghisellini1999}. The hard X-ray surveys (e.g. INTEGRAL, Swift) allow selection of extreme blazars \citep{ghisellini2009}. The group of known extreme HBL objects includes Mkn 501, 1ES 2344+514, PKS 0548-322 and 1ES 1426+428 \citep{costamante2001}, all of which are VHE sources. The two spectral components of extreme BL Lac objects peak at energies $\gtrsim 1$ keV and $\gtrsim 100$ GeV for the synchrotron and IC component respectively. The HE spectrum in the 100 MeV--100 GeV domain is expected to be hard. The lack of an HE counterpart of HESS~J1943+213 is unusual. Of all the VHE-detected AGN, only four (all HBL objects) have not been detected at GeV energies by {\em Fermi} \citep{fermiagn}. The upper limit by {\em Fermi} is quite strict but a similar situation has been observed e.g., in PKS 0548-322 \citep{hess2010b}.

The {\em Fermi}/LAT observations of TeV-selected AGN \citep{fermiTeV}, show that their HE spectra are hard with photon indices $\Gamma \lesssim 2.0$ where the extreme HBL objects belong to the hardest ones (i.e., 1ES 2344+514 with $\Gamma=1.57$ and 1ES 1426+428 $\Gamma=1.49$). Taking into account the {\em Fermi}/LAT upper limit the predicted photon index $\Gamma < 2$ of HESS~J1943+213 fits well within the HBL hypothesis.

Blazars are a class of objects known for their variability on all timescales. Nevertheless, in the duty cycle of VHE-emitting AGN, there are periods without signs of significant flux variability. Accordingly, the apparent lack of flux variability in case of HESS J1943+213 does not contradict the hypothesis of it being an HBL object. For instance, HBL object PG 1553+113, detected in both HE and VHE gamma rays, has shown only weak variability in radio, optical and X-rays and no variability in gamma-rays despite regular monitoring of the sky above 200 MeV by {\em Fermi}/LAT\citep{pg1553}. In addition, recent systematic studies of HE variability in blazars show that HBL objects are the least variable subgroup of blazars \citep{fermilightcurves}.

Based on the presented arguments, we conclude that the identification of HESS~J1943+213 as an extreme HBL object is the most plausible hypothesis.

\section{Conclusions}
\label{s:conclusions}
The \HESS array of Cherenkov telescopes has detected a new point-like VHE gamma-ray source located in the Galactic Plane. The source is not variable and has a soft spectrum. The VHE source is coincident with the unidentified INTEGRAL source IGR~J19443+2117, which was also detected in soft X-rays, radio and IR bands. The positional coincidence and lack of any other obvious counterpart suggest that these are the same object. The available archival data on IGR~J19443+2117 was combined with the new H.E.S.S., {\em Fermi} and radio observations. This allowed us to discuss the source classification as a gamma-ray binary, PWN and AGN, the three known types of sources that appear point-like in VHE observations.

The classification as HMXB appears less likely given the absence of a massive star. A faint flux of the infrared counterpart attributed to a massive star and assuming a similarity of the source SED with the known binaries, implies too large a distance ($>$25 kpc) and in consequence too high an X-ray luminosity. A secondary less constraining argument against the binary hypothesis is the absence of an orbital modulation.

The main argument against the PWN scenario is the soft VHE gamma-ray spectrum, which is unusual when compared with the spectra of a relatively large sample of TeV PWN. In addition, the unresolved X-ray emission as observed by {\em Chandra} weakens the PWN scenario.

There are no arguments against the AGN hypothesis. On the contrary, it is strongly supported by the radio, optical and X-ray properties being well within the parameter region (set by the broad band spectral indices and logarithmic flux ratios) in which high-frequency peaked BL Lac objects and TeV AGN are found. The infrared counterpart is compatible with an elliptical galaxy at $z>0.14$, which would also explain the soft VHE spectrum as attenuation of the intrinsic spectrum on the extragalactic background light. The apparent lack of lines in the preliminary infrared spectrum (if confirmed) could be explained since jet continuum emission usually dominates in this energy range. 

HBL objects comprise the majority of AGN detected until now in VHE gamma-rays. Statistically, assuming a uniform distribution on the sky, one out of 30 AGN can be expected to be situated at Galactic latitude $|b|<2\degr$. HESS~J1943+213 seems to be one of the most extreme HBL objects detected, with a peak frequency well beyond 10 keV and a low X-ray to radio ratio. 

The HBL classification would be secured by the identification of an elliptical galaxy in infrared images, the measurement of its redshift and the detection of correlated variability at radio, optical, X-ray or gamma-ray wavelengths. High resolution radio images from VLBI would also help to put further constraints on the source identity.


\begin{acknowledgements}
The support of the Namibian authorities and of the University of Namibia
in facilitating the construction and operation of H.E.S.S. is gratefully
acknowledged, as is the support by the German Ministry for Education and
Research (BMBF), the Max Planck Society, the French Ministry for Research,
the CNRS-IN2P3 and the Astroparticle Interdisciplinary Programme of the
CNRS, the U.K. Science and Technology Facilities Council (STFC),
the IPNP of the Charles University, the Polish Ministry of Science and 
Higher Education, the South African Department of
Science and Technology and National Research Foundation, and by the
University of Namibia. We appreciate the excellent work of the technical
support staff in Berlin, Durham, Hamburg, Heidelberg, Palaiseau, Paris,
Saclay, and in Namibia in the construction and operation of the
equipment.

The \textit{Fermi}/LAT Collaboration members acknowledge generous ongoing
support from a number of agencies and institutes that have supported both the
development and the operation of the LAT as well as scientific data analysis.
These include the National AFeronautics and Space Administration and the
Department of Energy in the United States, the Commissariat \`a l'Energie Atomique
and the Centre National de la Recherche Scientifique / Institut National de Physique
Nucl\'eaire et de Physique des Particules in France, the Agenzia Spaziale Italiana
and the Istituto Nazionale di Fisica Nucleare in Italy, the Ministry of Education,
Culture, Sports, Science and Technology (MEXT), High Energy Accelerator Research
Organization (KEK) and Japan Aerospace Exploration Agency (JAXA) in Japan, and
the K.~A.~Wallenberg Foundation, the Swedish Research Council and the
Swedish National Space Board in Sweden.

Additional support for science analysis during the operations phase is gratefully
acknowledged from the Istituto Nazionale di Astrofisica in Italy and the Centre National d'\'Etudes Spatiales in France.

 The authors thank T. Cheung, a referee on behalf of {\em Fermi}/LAT Collaboration, for valuable comments, N. Masetti for sharing the results of IR spectroscopy before publication and A. J. Bird at the University of Southampton and the IBIS survey team for providing the archival INTEGRAL/IBIS lightcurve of IGR J19443+2117.
\end{acknowledgements}

\newcommand{\thc}{(\hess\ Collaboration) }

\bibliographystyle{aa} 
\bibliography{aa16545} 

\end{document}